\documentclass[mathleft
]{an}
\usepackage{graphicx}

\usepackage{times}
\begin{document}

\Pagespan{1}{}
\Yearpublication{2010}%
\Yearsubmission{2010}%
\Month{07}%
\Volume{999}%
\Issue{88}%

\title{A search for magnetic fields in cool sdB stars\thanks{Based on observations obtained at the 
European Southern Observatory, 
Paranal, Chile (ESO programmes 083.D-0007(A) and 087.D-0049(A)).}}

\date{Received date / Accepted date}

\author{{G.~Mathys\inst{1,2}\fnmsep\thanks{Corresponding author: \email{gmathys@eso.org}}}
\and
S.~Hubrig\inst{3}
\and
E.~Mason\inst{4}
\and
G.~Michaud\inst{5}
\and
M.~Sch\"oller\inst{6}
\and
{F.~Wesemael\inst{5}\fnmsep\thanks{Deceased on September 28, 2011.}}
}

\titlerunning{Magnetic field in cool sdB stars}
\authorrunning{G.~Mathys et al.}

\institute{
Joint ALMA Observatory, Alonso de Cordova 3107, Vitacura, Santiago,
Chile
\and
European Southern Observatory, Casilla 19001, Santiago 19, Chile
\and
Leibniz-Institut f{\"u}r Astrophysik Potsdam (AIP), An der Sternwarte~16, 
14482~Potsdam, Germany
\and
Space Telescope Science Institute, 3700 San Martin Drive, Baltimore,
MD 21218, USA
\and
Depart\'ement de Physique, Universit\'e de Montr\'eal, Montr\'eal, PQ, H3C 3J7, Canada
\and
European Southern Observatory, Karl-Schwarzschild-Str.~2, 85748~Garching, Germany
}



\abstract{
Hot cluster Horizontal Branch (HB) stars and field subdwarf B
(sdB) stars are core helium burning stars that exhibit abundance
anomalies that are believed to be due to atomic diffusion.
Diffusion can be effective in these stars because they are slowly
rotating. In particular, the slow rotation of the hot HB stars
($T_{\rm eff} > 11,000$\,K), which show abundance anomalies, contrasts with
the fast rotation of the cool HB stars, where the observed
abundances are consistent with those of red giants belonging to
the same cluster. The reason why sdB stars and hot HB stars are
rotating slowly is unknown. In order to assess the possible role of
magnetic fields on abundances and rotation, we investigated the
occurrence of such fields in sdB
stars with $T_{\rm eff} < 30,000$\,K, whose temperatures overlap with
those of the hot HB stars. We conclude that large-scale organised
magnetic fields of kG order are not generally present in these stars
but at the achieved accuracy, the possibility that they have fields of
a few hundred Gauss remains open. We report the marginal detection of
such a field in SB\,290; further observations are needed to confirm
it. 
}

\maketitle

\section{Introduction}
\label{sect:intro}

Horizontal branch (HB) stars of clusters and their field equivalents, the subdwarf
B and O (sdB and sdO) stars, represent a challenge for stellar evolution theory. While they are all known to
burn He in their centre, the evolutionary scenario leading to sdB and sdO stars is not well established. It is not
well understood either why there are such relative differences as observed in the number of blue or extremely
blue HB stars from one cluster to the other. While metallicity is part of the explanation, it cannot account for
the diversity of HB branch morphologies observed. Reviews of typical characteristics
of HB stars in clusters are presented in the works of  Moehler (\cite{Moehler2001}) and Moehler \& Sweigart
(\cite{Moehler2006}).
In those globular clusters where HB stars with $T_{\rm eff} > 11,000$\,K are present, they have been observed to have
180metal abundances very different from those of the red giant (RG) branch stars of the same clusters, while the
HB stars with $T_{\rm eff} < 11,000$\,K have the same abundances as RG branch stars. For instance, 6 of the 7 HB stars
of M15 with $T_{\rm eff} > 11,000$\,K observed by Behr et al.\ (\cite{Behr1999}) (see their Fig. [1]) have [Fe/H] larger by a factor
of 50 than all cooler HB stars and the RG branch stars. Similar results were obtained in many other clusters
(Behr et al.\ \cite{Behr2000a}; Behr \cite{Behr2003}; Moehler et al.\ \cite{Moehler2000}; Fabbian et al.\ \cite{Fabbian2005}; Pace et al.\ \cite{Pace2006}). Furthermore,
the hotter HB stars rotate more slowly than the cooler ones that show no abundance anomalies (Behr et al.\ \cite{Behr2000a,Behr2000b}; Recio-Blanco et al.\ \cite{RecioBlanco2002})

The abundance anomalies in those hot HB stars are believed to be caused by atomic diffusion, radiative accelerations
leading for instance to the observed Fe overabundances (Michaud et al.\ \cite{Michaud2008}). The link with atomic
diffusion is strengthened by the observed slow rotation, a feature which is also a characteristic of Ap and HgMn
stars and is required to allow the slow diffusion processes to be effective. Similar statements can be made for the
sdB stars, for which both abundance anomalies of heavy elements
associated with diffusion processes (Geier et al.\ \cite{Geier2010}; Michaud et al.\ \cite{Michaud2011}) and very
slow rotation velocities are usually the norm (Edelmann \cite{Edelmann2003}). However the reason why very blue HB and sdB
stars rotate slowly is not known. Some suggestions have been made but are not generally accepted. Magnetic
fields are believed to be responsible for the slow rotation of Ap stars, but the origin of the slow rotation of HgMn
stars is also unknown. Could the presence of a magnetic field differentiate slowly rotating blue HB stars with
$T_{\rm eff} > 11,000$\,K and abundance anomalies from the red HB stars with the same composition as the RG stars?
Most sdBs and sdOs are also believed to have abundance anomalies caused by atomic diffusion. Gravitational
settling of He is important in most of them except for some of the sdOs where, in some cases, He is more
abundant than H.

\begin{figure}
\includegraphics[width=0.45\textwidth]{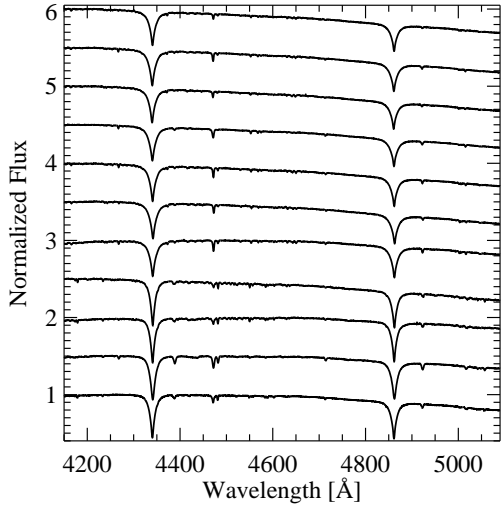}
\caption{
Stokes $I$ spectra of the ten sdB stars and Feige\,86 in the spectral region from 4200 to 5000\AA{}. From the bottom to the top 
we present: Feige 86, EC\,15327-1341, EC\,19490-7708, EC\,19579-4259, GD\,1110, LB\,1516, LB\,1559, SB\,290, 
SB\,410, SB\,459, and SB\,815. Note the similarity between most sdB spectra and the spectrum of Feige\,86.  
}
\label{fig:a}
\end{figure}

 In the past, magnetic fields of up to 1450\,G were detected at
significance levels ranging from 4 to 12\,$\sigma$ in four sdB and two sdO stars by O'Toole et al.\ (\cite{OToole2005}). A variable magnetic
field (from $\sim$0 to 10\,kG) was observed by Valyavin et al.\ (\cite{Valyavin2006}) in the sdO star Feige 34. Magnetic fields
of $-$1680\,G in an sdO star and varying between $-$1300 and 1750\,G in the sdOB star Feige\,66 were observed
by Elkin (\cite{Elkin1996}). Magnetic fields have now been observed in all sdO and sdB stars in which they were looked
for with an error bar lower than 1000\,G. Only upper limits had been observed for Feige\,86 and one sdO by
Borra et al.\ (\cite{Borra1983}) but the error bars are compatible with the more recent observations. All these stars have
effective temperatures close to or greater than 30,000\,K. It then appears likely that magnetic fields of kG order
are present in most if not all sdO and hot sdB stars. To our knowledge, there has been no attempt so far to
detect magnetic fields in cool sdBs or in HB stars of globular clusters.
While sdOs, sdBs and hot HB stars probably do not all have exactly the same evolutionary scenario, the fact
that they all burn He in their centre and the presence of diffusion-caused anomalies suggests that they are
strongly linked. It seems plausible that magnetic fields may be an important factor for all of them. Establishing
the presence or absence of such fields will provide important clues about their potential role in
slowing these stars down. This could for instance be done though magnetic fields forcing iso-rotation during their
preceding evolution, as magnetic fields have been suggested to do for the solar radiative interior (Charbonneau
\& MacGregor \cite{Charbonneau1993}).
To test the hypothesis that the observed abundance anomalies of sdOs, sdBs and hot HB stars
reflect the presence of magnetic fields in these stars, we conducted a
systematic search for magnetic fields in field sdB stars with $T_{\rm eff} < 30,000$\,K. 

\begin{figure}
\includegraphics[width=0.35\textwidth,angle=270]{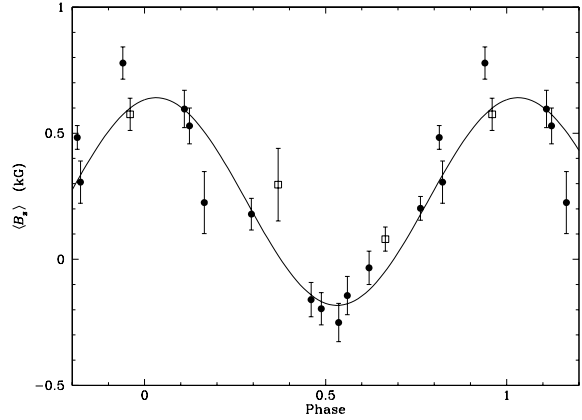}
\caption{
The FORS\,2 measurements of the mean longitudinal magnetic field of
the Ap star HD\,142070 that are reported here ({\em open squares\/})
are plotted together with the CASPEC measurements of Mathys et al.\ (in
preparation; {\em dots\/}) against phase, assuming a rotation period
of $3\fd37211$ and a phase origin ${\rm MJD_0=49877.7}$. The solid
curve is the best fit of the data by a sinusoid. 
}
\label{fig:hd142070}
\end{figure}
 
\section{Observations and magnetic field measurements}
\label{sect:obs}

We obtained FORS\,2 longitudinal magnetic field measurements
of ten sdB stars over three consecutive nights, from 28 to 31 August 2009.
Measurements of each star were based on a sequence of
observations with the following positions of the retarder
waveplate: $+$45, $-$45, $+$45, $-$45, etc. We used grism 600B and a slit width of  0$\farcs$4 to 
achieve a spectral resolving power of $R\approx2000$.  The observations were performed
using the readout mode (100kHz,high,1x1).
Most of the stars in our
sample were observed on each night to check the magnetic field
variability.  More details on the observing technique with FORS\,1 can be 
found elsewhere (e.g.,  
Hubrig et al.\ \cite{Hubrig2004a}, \cite{Hubrig2004b}, and references therein).
The mean longitudinal 
magnetic field, $\left< B_{\rm z}\right>$, was derived using 

\begin{equation} 
\frac{V}{I} = -\frac{g_{\rm eff} e \lambda^2}{4\pi{}m_ec^2}\ \frac{1}{I}\ 
\frac{{\rm d}I}{{\rm d}\lambda} \left<B_{\rm z}\right>, 
\label{eqn:one} 
\end{equation} 

\noindent 
where $V$ is the Stokes parameter that measures the circular polarisation, $I$ 
is the intensity in the unpolarised spectrum, $g_{\rm eff}$ is the effective 
Land\'e factor, $e$ is the electron charge, $\lambda$ is the wavelength, $m_e$ the 
electron mass, $c$ the speed of light, and ${{\rm d}I/{\rm d}\lambda}$ is the 
derivative of Stokes $I$.

\begin{table*}
\caption{Longitudinal magnetic field measurements of ten sdB stars.}
\label{tab:mess}
\centering
\begin{tabular}{lcr@{$\pm$}rrrr@{$\pm$}rrr}
\hline
\hline\\[-7pt]
\multicolumn{1}{c}{Object name} &
\multicolumn{1}{c}{MJD} &
\multicolumn{2}{c}{$\left<B_{\rm z}\right>^{\rm all}$} &
\multicolumn{1}{c}{$\overline{\left<B_{\rm z}\right>^{\rm all}} $} &
\multicolumn{1}{c}{$(\chi^2/n)^{\rm all}$} &
\multicolumn{2}{c}{$\left<B_{\rm z}\right>^{\rm hyd}$} &
\multicolumn{1}{c}{$\overline{\left<B_{\rm z}\right>^{\rm hyd}} $} &
\multicolumn{1}{c}{$(\chi^2/n)^{\rm hyd}$} \\
\multicolumn{1}{c}{} &
\multicolumn{1}{c}{} &
\multicolumn{2}{c}{[G]} &
\multicolumn{1}{c}{[G]} &
\multicolumn{1}{c}{} &
\multicolumn{2}{c}{[G]} &
\multicolumn{1}{c}{[G]} &
\multicolumn{1}{c}{} \\
\hline
HD\,142070 & 55071.9949 &    296 & 144 & 376 & 29.2 &    311 & 228 & 312 & 6.0 \\
HD\,142070 & 55072.9935 &     80 & 48  &     &      &    104 & 90  \\
HD\,142070 & 55073.9859 &    575 & 64  &     &      &    431 & 112 \\
EC15327    & 55072.0182 &    216 & 152 & 136 &  0.9 &    232 & 166 & 140 & 0.7 \\
EC15327    & 55073.0209 &     57 & 124 &     &      &     65 & 142 \\
EC15327    & 55074.0071 &  $-$76 & 128 &     &      &  $-$31 & 146 \\
EC19490    & 55072.0864 & $-$118 & 132 & 103 &  0.6 & $-$124 & 144 & 140 & 0.9 \\
EC19490    & 55073.0881 &    112 & 125 &     &      &    162 & 140 \\
EC19490    & 55074.0776 &  $-$77 & 147 &     &      & $-$133 & 152 \\
EC19579    & 55072.0443 &    322 & 120 & 215 &  3.3 &    317 & 133 & 226 & 3.0 \\
EC19579    & 55073.0485 &     30 & 65  &     &      &     32 & 124 \\
EC19579    & 55074.0300 &    188 & 118 &     &      &    229 & 126 \\
GD1110     & 55072.2234 &    279 & 180 & 249 &  1.7 &    294 & 205 & 355 & 2.7 \\
GD1110     & 55073.1538 &    216 & 211 &     &      &    408 & 220 \\
LB1516     & 55072.1442 & $-$370 & 182 & 402 &  4.8 & $-$373 & 212 & 455 & 4.8 \\
LB1516     & 55073.2173 & $-$433 & 185 &     &      & $-$526 & 206 \\
LB1559     & 55072.2705 & $-$159 & 202 & 224 &  1.2 & $-$256 & 220 & 298 & 1.8 \\
LB1559     & 55073.3097 &     63 & 178 &     &      &     80 & 202 \\
LB1559     & 55074.3159 & $-$349 & 208 &     &      & $-$441 & 221 \\
SB290      & 55072.3814 &     56 & 200 & 389 &  4.7 &    118 & 208 & 549 & 8.1 \\
SB290      & 55073.2561 & $-$639 & 188 &     &      & $-$856 & 209 \\
SB290      & 55074.2236 & $-$211 & 135 &     &      & $-$401 & 148 \\
SB410      & 55072.3089 & $-$457 & 182 & 339 &  3.5 & $-$508 & 210 & 369 & 3.1 \\
SB410      & 55073.3526 & $-$149 & 184 &     &      & $-$122 & 208 \\
SB459      & 55072.3478 &     48 & 180 & 290 &  2.1 &    144 & 198 & 321 & 2.1 \\
SB459      & 55073.3965 & $-$453 & 204 &     &      & $-$486 & 230 \\
SB459      & 55074.3839 &    215 & 188 &     &      &    229 & 192 \\
SB815      & 55072.1807 &  $-$97 & 204 & 261 &  1.6 & $-$204 & 212 & 352 & 2.3 \\
SB815      & 55072.4024 &    417 & 208 &     &      &    563 & 242 \\
SB815      & 55073.2783 & $-$178 & 184 &     &      & $-$216 & 208 \\
SB815      & 55074.2745 &    242 & 210 &     &      &    301 & 224 \\
\hline
\end{tabular}
\end{table*}

The longitudinal magnetic field was measured in two ways: using only the absorption hydrogen Balmer 
lines or using the entire spectrum including all available absorption lines.
In Fig.~\ref{fig:a} we present FORS\,2
integral spectra for all observed targets together with the well-studied
Pop\,II halo B-type star Feige\,86 with $T_{\rm eff} = 16\,430$\,K, in
which we searched for a magnetic field in May
2011. As we mention above, only upper limits have been
obtained for Feige\,86 by Borra et al.\ (\cite{Borra1983}). It is
however possible that Feige\,86 exhibits more similarity with  HgMn stars than with
sdB stars, as it shows He and Hg isotopic anomalies (Hubrig
et al.\ \cite{Hubrig2009}).

Our measurements of magnetic fields in ten sdB stars together
with the observations of the classical Ap star HD\,142070 (used as
a standard star) are shown in Table~\ref{tab:mess}. The first two columns list the object name and 
the modified Julian date of mid-exposure, 
followed by the measured longitudinal magnetic field $\left<B_{\rm z}\right>^{\rm all}$ 
using the whole spectrum. In columns 4 and 5 we give the rms field
$\overline{\left<B_{\rm z}\right>^{\rm all}}$ and the reduced
$\chi^2$. Columns 6 to 8 list $\left<B_{\rm z}\right>^{\rm hyd}$, 
$\overline{\left<B_{\rm z}\right>^{\rm hyd}}$, and  $(\chi^2/n)^{\rm hyd}$ for the measurements using 
hydrogen lines. 

In order to minimize the risk of apparent non-detection in some of the
targets of our sample, due to fortuitous null observations of the
longitudinal field close to the phases where it reverses its sign,
stars were observed at two to four different epochs. The rms field is defined as:
\begin{equation}
\overline{\left<B_{\rm z}\right>}=\left({1\over
    n}\sum_{i=1}^n\left<B_{\rm z}\right>_i^2\right)^{1/2}\,,
\end{equation}
and the reduced $\chi^2$ as
\begin{equation}
\chi^2/n={1\over n}\sum_{i=1}^n\left({\left<B_{\rm
        z}\right>_i\over\sigma_i}\right)^2\,,
\end{equation}
where $n$ is the number of measurements of the considered star,
$\left<B_{\rm z}\right>$ is the $i$-th such measurement and $\sigma_i$
  is its uncertainty.

Figure~\ref{fig:hd142070} shows the three measurements of the mean
longitudinal magnetic field of the Ap star HD\,142070 that we obtained
from consideration of its whole spectrum, together with the
measurements of Mathys et al. (in preparation), based on CASPEC
spectropolarimetric observations. The good
agreement between the two datasets confirms the quality of the
longitudinal field determinations achieved from FORS\,2 observations
and indicates that the order of magnitude of their quoted
uncertainties is correct.

\section{Discussion}
\label{sect:discussion}
In no star do our measurements reveal the presence of 1-2\,kG
fields. Our ability to detect weaker fields is limited by the accuracy
of the measurements, as a result of the faintness of the studied stars
and of the readout mode that was used. In only one case, SB\,290, the
longitudinal field determined at one of three epochs is formally
significant at a level greater than 3$\sigma$. For this star, the
reduced $\chi^2$ of the three measurements that were performed also
supports the reality of a detection at a confidence level greater than
99\%. However, this conclusion depends critically on the correctness
of the adopted measurement uncertainty; if the latter was only
slightly underestimated (by 20\% for the measurements based on all
absorption lines), it would be invalidated. Thus measurements at more
epochs and with better accuracy are needed to confirm the presence of
a magnetic field in SB\,290. 

On the other hand, no significant longitudinal field was detected in Feige\,86 in
our recent spectropolarimetric observations of May 2011. The
measurements resulted in $\left< B_z\right>_{\rm all} = 55\pm49$\,G.

In conclusion, this study shows that large-scale organised magnetic
fields of kG order are not generally present in sdB stars with $T_{\rm
  eff}<30\,000$\,K. Yet it leaves open the possibility that these
stars may have fields of a few hundred Gauss, with in particular a
tantalising, although marginal, detection in one of them, SB\,290. A
firmer conclusion will require additional observations of higher
quality.  



\label{lastpage}

\end{document}